\documentclass[twocolumn,showpacs,amsmath,amssymb,floatfix]{revtex4}
\usepackage{graphicx}
\usepackage{hyperref} 
\usepackage{latexsym}

\begin{document}  
  
\def\av#1{\langle #1 \rangle}
\def\N#1{N^{(#1)}}
\def\n#1{n^{(#1)}}
\def\l{\lambda}

\title{Designer Nets from Local Strategies}  

\author{Hern\'an D. Rozenfeld}\email{rozenfhd@clarkson.edu}
\author{Daniel ben-Avraham}\email{benAvraham@clarkson.edu}
\affiliation{Department of Physics, Clarkson University, Potsdam NY 13699-5820}

\date{\today}

\begin{abstract}  
We propose a {\em local} strategy for constructing scale-free networks of arbitrary degree distributions,
based on the redirection method of Krapivsky and Redner [{\it Phys. Rev. E} {\bf63}, 066123 (2001)].  Our method includes a set of external parameters that can be tuned at will to match detailed behavior at small degree $k$, in addition to the scale-free power-law tail signature at large $k$.  The choice of parameters determines other network characteristics, such as the degree of clustering.  The method is local in that addition of a new node requires knowledge of only the immediate environs of the (randomly selected) node to which it is attached.  (Global strategies require information on finite fractions of the growing net.)  
\end{abstract}  

\maketitle

Recently much effort has been devoted to the study of large networks that surround us in everyday life, such as the Internet and the World Wide Web, the electricity power grid, flight connections, social networks of contacts or collaborations, networks of predator-prey, of neurons in the brain, 
etc.,~\cite{albert02,dorogovtsev02,bornholdt}.  An important realization is that a majority of these networks share some characteristic properties: A small diameter --- a small number of links connects between any two nodes on the net; clustering, or the small world property --- the neighbors of a node tend to be connected to one another; and a scale-free degree distribution --- the distribution of the number of links emanating from a node (the degree $k$) has a power-law tail of the form 
\begin{equation}
P(k)\sim k^{-\l}.
\end{equation}
The scale-free property gives rise to exotic behavior of the networks, such as resilience to random dilution (the percolation transition does not take place for $\l<3$), on the one hand, and high vulnerability to removal of the most connected nodes, on the other hand, and has become a principal focus of attention~\cite{resilience,attack}.
  
Several growth models that produce scale-free networks have been suggested.  However, most
growth techniques rely on {\em global} properties of the network.  Such is the case, for example, for the seminal model of Barab\'asi and  Albert  (BA)~\cite{internet}, where new nodes are connected to an existing node with a probability proportional to its degree.  The BA algorithm requires global knowledge of the degree of all present nodes.  While global algorithms have contributed immensely to our understanding of how scale-free degree distributions might emerge, in most common situations it is more likely that networks evolve by a set of {\em local} rules. (One does not typically conduct a survey of the Internet for deciding where to connect a new router, neither does one study a whole network of social contacts for selecting new acquaintances.) 

Here we propose a {\em local} strategy for constructing scale-free networks, based on the redirection technique of Krapivsky and Redner (KR)~\cite{KR}.  Our method includes a set of external parameters that enable fine tuning of additional properties (degree distribution at small $k$, degree of clustering, etc.), while guaranteeing a scale-free tail with a given degree distribution exponent, as in the original KR method. 

\begin{figure}
\includegraphics*[width=0.25\textwidth]{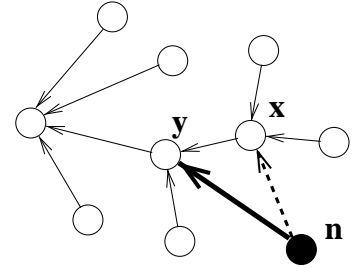}
\caption{The KR redirection model.  A new node {\bf n} (solid) is connected to a randomly selected
node {\bf x} with probability $1-r$ (dashed arrow), or the link is redirected to node {\bf y}, the ancestor of {\bf x}, with probability $r$ (solid thick arrow).  After Krapivsky and Redner~\cite{KR}.
}
\label{R}
\end{figure}

%
In the KR redirection model, a new node {\bf n} is connected to a pre-existing node {\bf x}, selected at random ({\em without} regard to its degree), with probability $1-r$.  If the connection is established, {\bf x} is said to be the {\em ancestor} of {\bf n} (the link may be regarded as directed, from {\bf n} to {\bf x}).  With probability $r$, the connection is not realized, but instead it is {\em redirected} to {\bf y}, the ancestor of {\bf x} (Fig.~\ref{R}).

The redirection step is essentially what turns the global BA strategy of selecting a node with a probability proportional to its degree, to a local one.  (See~\cite{feld}, for an early introduction of this basic principle.) Each node has but one ancestor, so a node of degree $k$ has $k-1$ incoming links.  It follows that an ancestor of degree $k$ is reached with a probability proportional to $k-1$, but purely through a local process.  The KR technique builds scale-free graphs with degree exponent $\l=1+1/r$ (that is, $2\leq\l\leq\infty$).  The BA net corresponds to the special case of $r=1/2$ ($\l=3$).  However, the obtained graphs are always {\em trees} --- they possess no cycles (a unique path connects between any two nodes) --- and not much else can be controlled during the growth process, beyond the degree distribution exponent.

We generalize the redirection algorithm as follows.  At each time step the new node {\bf n} is connected to $m$ different pre-existing nodes, with probability $p_m$, $m=1,2,\dots$, $(\sum_mp_m=1)$. Each of the $m$ nodes is selected randomly and independently by the same redirection trick as for a single node: direct connection to the selected node with probability $1-r$, or redirection to a random ancestor, with probability $r$.  The KR model corresponds to the choice $p_m=\delta_{m,1}$.  The presence of an initial network seed, of $N_0$ nodes, is implied.  The seed allows for the process to get started.

A new node is added at each time step, so the total number of nodes increases as
\begin{equation}
N(t)=N_0+t.
\end{equation}
In the long time asymptotic limit of $t\gg N_0$, $N(t)\sim t$.
Because each new node has $m$ ancestors with probability $p_m$, the average number of ancestors
of a node at time $t$ is
\begin{equation}
\av{m}(t)=\frac{N_0\av{m}_0+t\sum mp_m}{N_0+t},
\end{equation}
where $\av{m}_0$ is the initial average number of ancestors present in the seed.  In the long time asymptotic limit, $\av{m}\sim\sum mp_m\equiv a$.  The average degree of a node (including both incoming and outgoing links) is 
\begin{equation}
\av{k}=2\av{m}\sim 2a.
\end{equation}

Consider now $\N{l}_k(t)$, the number of nodes of degree $k$ that possess exactly $l$ ancestors. It  evolves according to
\begin{eqnarray}
\label{dNk/dt}
&&\frac{d\N{l}_{k}(t)}{dt} = \sum_m mp_m\{\frac{(1-r)}{M_0} [\N{l}_{k-1}-\N{l}_{k}] \nonumber\\
&&+\frac{r}{M}[(k-l-1)\N{l}_{k-1} - (k-l)\N{l}_{k}]\}+p_{l}\delta_{k,l},\quad k\geq l;\nonumber\\
&& \N{l}_k=0,\qquad k<l.
\end{eqnarray}
The first term inside the curly brackets describes gains and losses due to direct connections (occurring with probability $1-r$), while the additional terms refer to changes due to redirection: since the nodes in question have $l$ ancestors, the probability to reach a node of degree $k$ by redirection is now proportional to $k-l$, instead of $k-1$, as in the original KR model. Note that the number of ancestors of a node is fixed at birth, and does not change subsequently.  The rate for directed connections is normalized by $M_0=\sum_l\sum_k\N{l}_k=N\sim t$, while for redirected events, 
\begin{equation}
\label{M}
M=\sum_l\sum_k(k-l)\N{l}_k=N\big(\av{k}-\sum_l\frac{\N{l}}{N}l\big).
\end{equation}
In the long time asymptotic limit $\N{l}/N\to p_l$, hence $M\sim N(\av{k}-a)\sim at$~\cite{remark}.

Due to the time dependence of the rates in Eq.~(\ref{dNk/dt}), analytic integration is not feasible.  Instead, we argue that the transient time does not affect the ultimate
distribution appreciably, and replace $M_0$  and $M$ by their long time asymptotic limits.  
From here we proceed as in~\cite{KR}.  The asymptotic equations admit the general solution
$\N{l}_k(t)=\n{l}_kt$, where
\begin{eqnarray}
\label{nkl_recursion}
&&\n{l}_{k}= a(1-r)[\n{l}_{k-1}-\n{l}_{k}] \nonumber\\
&&+r[(k-l-1)\n{l}_{k-1} - (k-l)\n{l}_{k}]\}+p_{l}\delta_{k,l},\quad k\geq l;\nonumber\\
&& \n{l}_k=0,\qquad k<l.
\end{eqnarray}
Note that $\n{l}_k$ is in fact normalized (since $N\sim t$) and represents the {\it probability} that a randomly selected node be of degree $k$ and have $l$ ancestors.

The solution of the recursions (\ref{nkl_recursion}) is  $\n{l}_k=\Pi_{k'}[1-(1+r)/(rk'+1+a(1-r)-rl)]$.  Writing the product as the exponential of a sum, 
expanding to first order in $1/k'$ (for large $k'$), and approximating the sum by an integral, one
finally obtains 
\begin{equation}
\n{l}_k\sim k^{-(1+1/r)}.
\end{equation}
The same is true for the overall degree distribution (without reference to the number of ancestors), \begin{equation}
P(k)=n_k=\sum_{l=1}^k\n{l}_k\sim k^{-(1+1/r)}.
\label{Pk-nk}
\end{equation}
This follows from Eqs.~(\ref{nkl_recursion}), summing over $l$, noting that
$\sum_l l\n{l}_k=a$, and proceeding as for $\n{l}_k$.

We now have at our disposal a simple local algorithm that produces scale-free graphs of any desired degree exponent $\l=1+1/r$, as well as a set of parameters (the $\{p_m\}$) that can be tweaked at will to achieve a wide gamut of additional attributes, without affecting $\l$.  Obvious examples include the average number of outgoing links from a node, $\av{m}=\sum mp_m=a$, and the average node degree, $\av{k}=2a$.  (Note that in the original KR model $a=1$ is the only possible
outcome.)

\begin{figure}
\includegraphics*[width=0.4\textwidth]{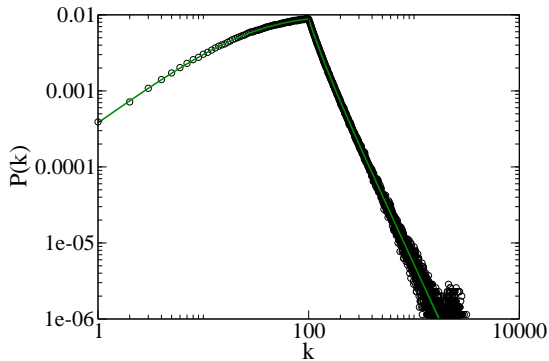}
\caption{Degree distribution of network constructed with $p_m=1/100$, $m=1,2,\dots,100$, and $r=0.5$ ($\circ$).  The analytical prediction from Eqs.~(\ref{nkl_recursion}) and (\ref{Pk-nk}) (solid line) is shown for comparison.}
\label{p100} 
\end{figure}

The degree distribution at small $k$ is determined by the particular choice of the $\{p_m\}$.  This is dramatically illustrated  in Fig.~\ref{p100}, where we show the degree distribution of a network grown with $r=1/2$ and $p_m=1/100$ for $m=1,2,\dots,100$ ($p_m=0$ for $m>100$).  In addition to the expected $k^{-3}$ tail, we see a very distinct distribution, for $k<100$.   The resulting distribution is perfectly reproduced, analytically, using Eqs.~(\ref{nkl_recursion}).

More importantly, the inverse problem is easily solved as well.  Given a specific distribution $P(k)$, with a scale-free tail $\sim k^{-\l}$, it can be produced by the redirection method in the following way.
Set $r=1/(\l-1)$ for the redirection probability, thus ensuring the desired scale-free tail.  Compute
$a=(1/2)\sum kP(k)$ that is implied by the desired distribution.  According to Eq.~(\ref{nkl_recursion}), 
\begin{equation}
p_m=[1+a(1-r)]\n{m}_m.
\end{equation}
The required $\n{m}_m$ are computed iteratively, from Eqs.~(\ref{nkl_recursion}) and (\ref{Pk-nk}).
From (\ref{Pk-nk}) we have $\n{1}_1=P(1)$.  Using this, and Eqs.~(\ref{nkl_recursion}), for $k=2$, $l=1$,
one can compute $\n{1}_2$, then obtain $\n{2}_2=P(2)-\n{1}_2$ from (\ref{Pk-nk}), etc...

\begin{figure}
\includegraphics*[width=0.4\textwidth]{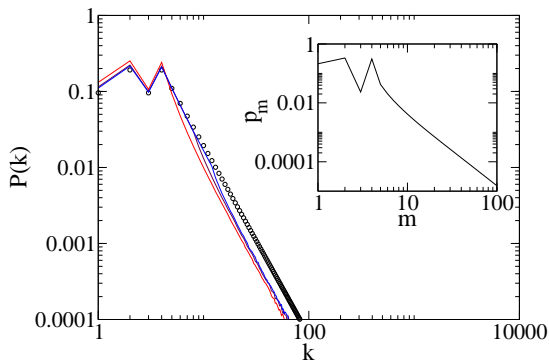}
\caption{Convergence to a target distribution ($\circ$) for nets grown with the analytically computed $p_m$'s (inset).  Shown are distributions obtained for networks grown with just the first $4$, $8$, and $12$ $p_m$'s (solid curves).  Convergence to the target distribution improves with increasing range of $m$. }
\label{inverse} 
\end{figure}

We have followed this procedure to compute the $p_m$ needed to grow a network with the target distribution
$P(k)=c,2c,c,2c$, for $k=1,2,3,4$, respectively, and $P(k)=2c(k/4)^{-2.5}$, for $k\geq 4$, ($c= 1/10.4359$, is determined from normalization:  $\sum P(k)=1$), shown in Fig.~\ref{inverse}.  As might be expected, the $p_m$ decay rapidly with $m$ (in this particular case, $p_m\sim m^{-2.5}$) and a reasonably well-fitting distribution is obtained with just $m\leq4$.
Increasing the range of $m$ yields a better fit to the target.   Alternatively, a better fit can be achieved, for a fixed range of $m$, by treating $a$ as a variable whose value is then optimized. 

Another interesting property is the number and distribution of loops, or  cycles, in the growing nets.  One measure of cycles is provided by the
{\em clustering index}, defined for site $i$ as~\cite{watts98}
\begin{equation}
C_i=\frac{E_i}{\frac{1}{2}k_i(k_i-1)}.
\end{equation}
It denotes the ratio between $E_i$, the actual number of links between the $k_i$ neighbors of the node, to the maximum number of links that would result had all the $k_i$ nodes been connected to
one another. 
For trees, the clustering index is zero at all nodes.  In contrast, the proposed generalized KR process can produce loops, and hence nonzero clustering,  whenever a new node is connected to more than
one site: $p_1<1$ (or $p_m>0$ for some $m>1$).

To test this issue we focused on the KR model, but where a newly added node is further connected to a second node with probability $p$: $p_1=1-p$, $p_2=p$  (and $p_m=0$ for $m>2$).
Several networks were grown in this way, with different values of $r$ and $p$.  The numerical
data strongly support the predicted relation $\l=1+1/r$ and the independence from the $p_m$.  

\begin{figure}
\includegraphics*[width=0.36\textwidth]{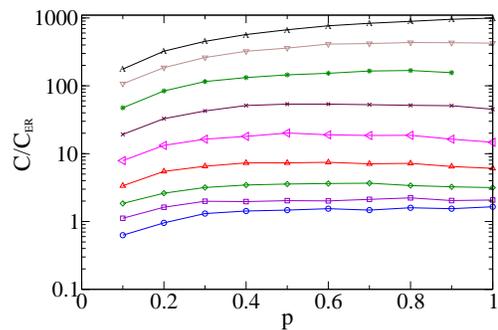}
\caption{Clustering index, $C$, for nets constructed in the redirection method, with $r=0.1,0.2,\dots,0.9$ (bottom to top) and $p_m=(1-p)\delta_{m,1}+p\delta_{m,2}$.  Plotted is $C/C_{\rm ER}$, where  $C_{\rm ER}$ is the clustering index of the equivalent ER graphs~\cite{albert02,remark2}.
}
\label{C/Cer}
\end{figure}

\begin{figure}
\includegraphics*[width=0.36\textwidth]{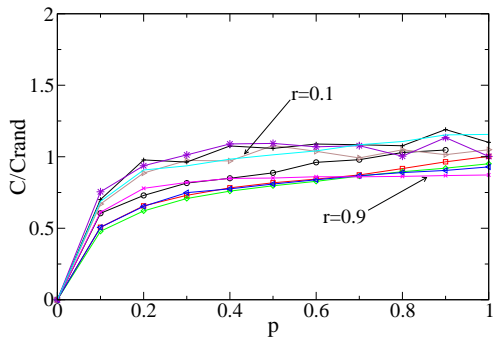}
\caption{Clustering index for the same data as in Fig.~\ref{C/Cer}, but compared to equivalent
scale-free random graphs (see text).}
\label{C/Crand}
\end{figure}

The results for clustering are summarized in Figs.~\ref{C/Cer} and \ref{C/Crand}.  In Fig.~\ref{C/Cer}, we plot
the ratio of the global cluster coefficient (averaged over all the sites of the network), $C$, to the cluster
coefficient of an equivalent Erd\H os-R\'eniy (ER) graph~\cite{albert02,remark2}, $C_{\rm ER}$, as a function of $p$, for various values of $r$.  As might be expected, the clustering index increases with $p$, though the effect saturates
rather quickly.  A more pronounced effect is achieved by increasing $r$, which changes $C$ by orders
of magnitude (note the logarithmic scale).  This has to do with the influence of $r$ on the degree
distribution.  To see that, in Fig.~\ref{C/Crand} we plot, for the same data, the ratio of $C$ to $C_{\rm rand}$ --- the clustering index in equivalent networks, with identical scale-free degree distributions, but where all the connections have been redistributed randomly.  The dependence on $r$ seems a lot weaker by this comparison.  Fig.~\ref{C/Crand} demonstrates, however, significant differences between random scale-free networks and nets constructed by the generalized redirection method.

\begin{figure}
\includegraphics*[width=0.4\textwidth]{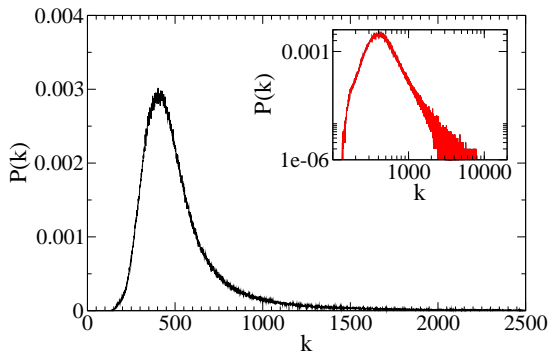}
\caption{Degree distribution of ``self-consistent" net grown with $p_m=P(m,t)$.  The distribution
peaks at a value of $k$ that increases with time, while its tail might be possibly interpreted as scale-free (inset):  $\l=1+1/r=3$ is consistent with our noisy data.}
\label{canon}
\end{figure}

As a final remark, we note that we have tacitly assumed that the set $\{p_m\}$ is finite.  Indeed, were the set infinite, there would be no way to even start the growth process, unless one had an infinitely large seed as well, since connecting a node to $m$ random nodes requires that there be at least $m$ nodes present to begin with.
One way to get around the limitation to finite sets, is by allowing the $p_m$ to vary with time.  Consider, for example, the ``self-consistent" networks grown with $p_m(t)=P(m,t)$, that is, where the $p_m$ reflect the existing degree distribution at time $t$.  The degree distribution that results from this strategy, using $r=1/2$, is shown in Fig.~\ref{canon}.  We have not attempted analyzing this distribution, because the equations involved are quite more complicated, and we cannot think of any serious applications to this curious model.  Still, it remains an amusing problem that might test the limits of the general approach of evolution equations.

We have presented a local strategy --- the generalized redirection method --- for the growth of scale-free networks.  Local algorithms require information about only the immediate environs of a visited node, and are more likely to reflect real growth processes than global algorithms, where addition of new nodes requires information about the whole network.  The generalized redirection technique includes a class of external parameters, the $p_m$, that may be tuned at will, without affecting the degree exponent (which is controlled by the independent parameter $r$), to achieve a variety of effects.  We have explicitly demonstrated how to design nets with arbitrary degree distributions, at small degree $k$, and power-law (scale-free) tails, at large $k$.  We have also shown that additional attributes, such as the degree of clustering, can be manipulated by a proper choice of the model's parameters.  However,
how to do this effectively remains an open question.  We anticipate that analogous algorithms to the one presented for the degree distribution can be designed for other attributes as well, by studying the relevant rate equations.

\acknowledgments

We thank Erik Bollt for useful discussions, and Sid Redner for helpful discussions and for providing us with Fig.~\ref{R}.  We are grateful to NSF Grant No.~PHY-0140094 for partial financial support of this work.



\begin{thebibliography}{99}

\bibitem{albert02} 
   R.~Albert and A.-L.~Barab\'asi, {\it Rev. Mod. Phys.} {\bf74}, 47 (2002).

\bibitem{dorogovtsev02} S.~N.~Dorogovtsev and J.~F.~F.~Mendes, {\it Adv. Phys.} {\bf51}, 1079 (2002).

\bibitem{bornholdt} S.~Bornholdt and H.~G.~Schuster, {\it Handbook of Graphs and Networks},
(Wiley-VCH, Berlin, 2003).   

\bibitem{resilience} R.~Albert, H.~Jeong, and A.-L.~Barab\'asi, {\it Nature} (London) {\bf406}, 378 (2000);
{\bf406}, 6794 (2000).  R.\ Cohen, K.\ Erez, D.\ ben-Avraham, and S.\ Havlin, {\it Phys. Rev. Lett.} {\bf85}, 4626 (2000).

\bibitem{attack} R.\ Cohen, K.\ Erez, D.\ ben-Avraham, and S.\ Havlin, {\it Phys. Rev. Lett.} {\bf86}, 3682 (2001).


\bibitem{internet}
   A. L. Barab\'asi, R. Albert,
   {\it Science}, {\bf 286}, 509 (1999).

\bibitem{KR} P.L. Krapivsky and S. Redner, {\it Phys. Rev. E} {\bf63}, 066123 (2001);
{\it J. Phys. A} {\bf35}, 9517 (2002); J. Kim, P.L. Krapivsky, B. Kahng, and S. Redner, {\it Phys. Rev. E} {\bf66}, 055101(R) (2002).

\bibitem{feld}   {\it Am. J. Sociology} {\bf 96}, 1464 (1991).
See, also:
   M. E. J. Newman,	
   {\it Social Networks} {\bf 25}, 83 (2003).  

\bibitem{remark}  In the KR model the normalization factor $M\to 1$, in the long time asymptotic limit, and can be justifiably neglected.

\bibitem{remark2} An equivalent ER graph is an Erd\H os-R\'eniy random graph with the same number of
nodes, $N$, and the same number of links, $M$, as in the original scale-free graph.  It is easy to show that $C_{\rm ER}=2M/N(N-1)$.  That is, in our case $C_{\rm ER}=2(1+p)/(N-1)$.  

\bibitem{watts98} D.~J.~Watts and S.~H.~Strogatz, {\it Nature} (London), {\bf393}, 440 (1998).

\end{thebibliography}
\end{document}